# Surface Binding Energies for Amorphous Plagioclase Feldspar Calculated using Molecular Dynamics

Amanda Ricketts[1,2], Benjamin A. Clouter-Gergen[2], Anastasis Georgiou[2], Deborah Berhanu[3,4], Liam S. Morrissey[2,3,4]

**Abstract**

Despite the well-established presence of amorphous compounds on planetary bodies such as the Moon and Mercury due to space weathering, the direct effect of atomic arrangement on the surface binding energies (SBEs) of elements on these bodies remains largely unexplored. Accurate SBE values are essential for reliably predicting sputtering yields and the energy distribution of ejecta. Here, we use molecular dynamics simulations to quantify SBEs for the different elements sputtered from amorphous atomic arrangements of the plagioclase feldspar end members, albite and anorthite, and compare to their crystalline counterparts. Results show that while the mean elemental SBEs from amorphous surfaces are not significantly different from their crystalline counterparts, the random orientation in amorphous structures gives rise to a spectrum of bonding configurations, resulting in a distribution of SBEs with a wider range. This contrasts with the clearly discretized set of SBE values associated with the ordered atomic structure of crystalline surfaces. We then consider sputtering by H, He, and a solar wind combination of 96% H and 4% He. For each of these cases, we demonstrate that there is minimal difference (<10% for albite and <20% for anorthite) between the sputtering yields of amorphous and crystalline surfaces. We attribute these results to the presence of the same elemental bonds across different atomic arrangements, which leads to similar mean SBEs and, consequently, comparable sputtering yields.

[1] Corresponding author apr643@mun.ca
[2] Memorial University of Newfoundland, St. John's, NL
[3] Fashion Institute of Technology, New York, NY 10001, USA
[4] American Museum of Natural History, New York, NY 10024, USA

# 1 INTRODUCTION

The properties of the surface and surrounding exosphere of airless bodies, such as Mercury and the Moon, are affected by the process of space weathering, the interaction of the planetary surface with solar wind ions, micrometeoroid impacts, photon stimulated desorption, and related processes (Hapke 2001; Bennett et al. 2013; Domingue et al. 2014; Pieters & Noble 2016). Our focus here is on the surface sputtering due to solar wind ions. The contribution of ion sputtering to the exosphere is known to be strongly dependent on the energy needed to remove an atom from the surface of the compound in which it is bound, a quantity known as the surface binding energy (SBE) (Kelly 1986; Behrisch & Eckstein 2007; Killen et al. 2022; Szabo et al. 2024; Brötzner et al. 2025). Recent studies on this SBE have used molecular dynamics (MD) modelling to simulate crystalline silicate surfaces directly on the atomic scale. These MD-derived SBEs for the different atom types were found to be approximately seven times higher than the commonly assumed monoelemental cohesive energies (Killen et al. 2022; Morrissey et al. 2022b, 2024). The differences in SBEs were attributed to the local atomic arrangement of the atom on the surface, along with the collection of partial bonds formed with neighboring atoms. These values were then used as input into binary collision approximation (BCA) and global exosphere models to show their effects on sputtering and exosphere composition. Lower SBE atoms were sputtered more efficiently and at lower energies, contributing to the exosphere at lower altitudes. In contrast, higher SBE atoms were retained longer and, when ejected, often exceeded the escape energy of the body.

While these studies demonstrated that SBEs are sensitive to both crystallographic orientation and local atomic environment, they were limited to minerals with crystalline atomic arrangements only. However, as these surfaces are exposed to the space weather, they are expected to quickly develop damage within a near surface rim. Atoms in this rim lose the long-range order found in crystalline materials and are thus often considered to be "amorphous" (Grossman et al. 1970; Keller & McKay 1991, 1994, 1997). Given that SBEs are known to be a function of the atomic arrangement on the surface, it follows that amorphous substrates likely have SBEs distinct from their crystalline counterparts. However, no study has quantified the SBEs for elements from amorphous minerals and their effect on the predicted sputtering yield. Studying amorphous materials alongside their crystalline counterparts is crucial. Here, we have used MD to calculate the SBE of all element types within amorphous albite ($NaAlSi_3O_8$) and anorthite ($CaAl_2Si_2O_8$), the two endmembers of the plagioclase feldspars.

## 2 METHODOLOGY

### 2.1 MD Simulations

MD simulations were performed using the Large-scale Atomic/Molecular Massively Parallel Simulation (LAMMPS) package (Plimpton 1995; Thompson et al. 2022). A reactive force field (ReaxFF) empirical potential was used to simulate interactions between atoms in each mineral, allowing for the dynamic simulation of bond breaking and reformation in a multielemental substrate (Van Duin et al. 2001). ReaxFF efficiently models both bonded and nonbonded interactions. It accounts for connectivity-dependent reactions, setting the energy contribution of broken bonds to zero, while nonbonded interactions (van der Waals and Coulomb forces) are calculated independently of bonding. Further details can be found in Van Duin et al. (2001). For this study, a ReaxFF potential originally developed by Pitman & Van Duin (2012) to accurately simulate aluminosilicates (i.e., materials that include Al, Si, O, Na, and H) was used. Not only was this potential specifically parameterized for silicates and Na-O interactions, but it has also been widely applied to model various silicates, showing good agreement with mineral densities. It has further been used to quantify the SBEs of crystalline albite and anorthite (Lyngdoh et al. 2019; Mayanovic et al. 2023; Morrissey et al. 2024). Moreover, Yu et al. (2017) evaluated the potential for its ability to reproduce different amorphous silicate configurations, highlighting its applicability across diverse compositions and atomic arrangements.

#### 2.1.1. Preparing Mineral Substrates

We first used MD to develop and equilibrate amorphous substrates containing approximately 35000-37000 atoms, allowing for ~150 surface atoms per element type to sample the SBE distribution. We obtained the mineral's conventional unit cell from the Materials Project database (Jain et al. 2013) and replicated to achieve substrates that were ~150Å x 150Å x 25Å in the x, y, z dimensions, respectively. To form the amorphous substrates, we then followed an MD melting/cooling procedure that has been used previously on silicates (Fogarty et al. 2010; Morrissey et al. 2022a; Pallini et al. 2023). This procedure begins with a crystalline mineral and ultimately achieves an amorphous substrate without long-range order, likely representing the amorphous rims formed on planetary surfaces. This involved first heating the entire system to 4000K, holding at this temperature for 75ps, and then cooling the system back to 1K using a Nosé-Hoover thermostat with an NVT (number of particles, volume, temperature) ensemble that keeps the volume and number of particles fixed while allowing the temperature to change. Next, the system was again heated to 4000K, held at this temperature for 75ps, and cooled back to 1K, this time using an isobaric-isothermal NPT (number of particles, pressure, temperature) ensemble that fixed the number of particles while allowing the volume to change based on the desired temperature and pressure. This heat treatment process resulted in an amorphous albite density of 2.21g/cm$^3$ and an amorphous anorthite density of 2.51g/cm$^3$, agreeing well with experimental results and validating the approach (Dal Bó et al. 2013; Cui et al. 2019). Next, the simulation domain was extended 100Å in the z direction to open the surface of interest to vacuum. To simulate an infinite slab with constant thickness and a free surface, the boundary conditions were set to fixed in the z direction and periodic in the x and y directions. Atoms in the bottom three angstroms in the fixed direction of each substrate were fixed in space to ensure there is only one surface and no effects from bottom atoms, a standard practice for slab simulations (Yang & Hassanein 2014; Morrissey et al. 2024). The resulting amorphous albite and anorthite surfaces are shown in Figure 1.

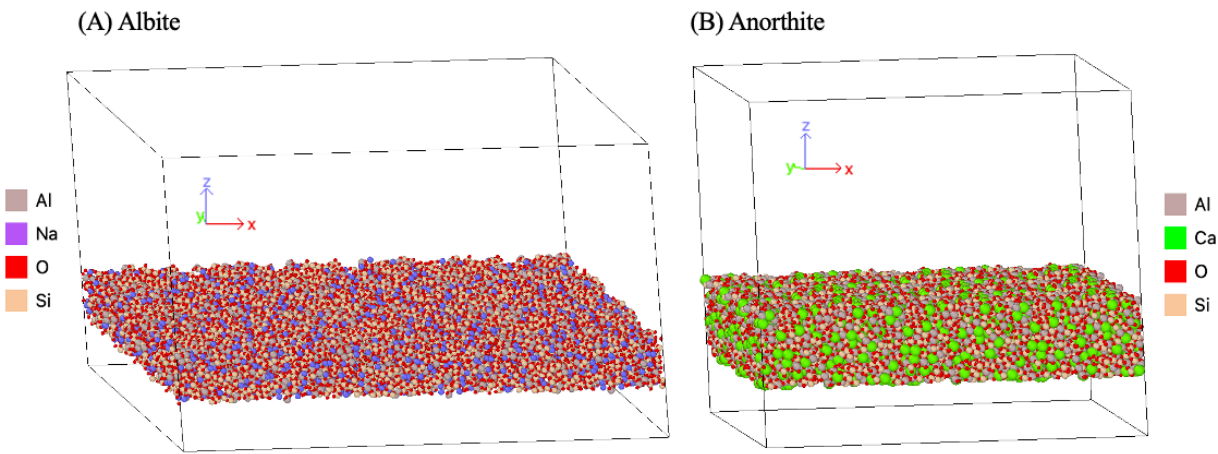

**Figure 1** Albite (A) and anorthite (B) MD substrate surfaces.

### 2.1.2. Determining the Surface Atoms

A process was established to differentiate the surface atoms from the bulk atoms in the substrate using OVITO (Stukowski 2010), an open-source visualization and analysis software for atomistic and particle simulation data. In OVITO, a probe sphere with a radius of 3.4Å was created to sample surface atoms. This radius was small enough to "roll" over the outer surface, but too big to probe into internal spaces, identifying only the atoms exposed to the vacuum. To isolate only the topmost surface atoms, an expression selection was then applied to include atoms within approximately 3Å below the topmost atom.

This range effectively captured the irregular topography of the amorphous surfaces while minimizing the inclusion of underlying, non-surface atoms. Figure 2 shows the resulting surface atoms in OVITO.

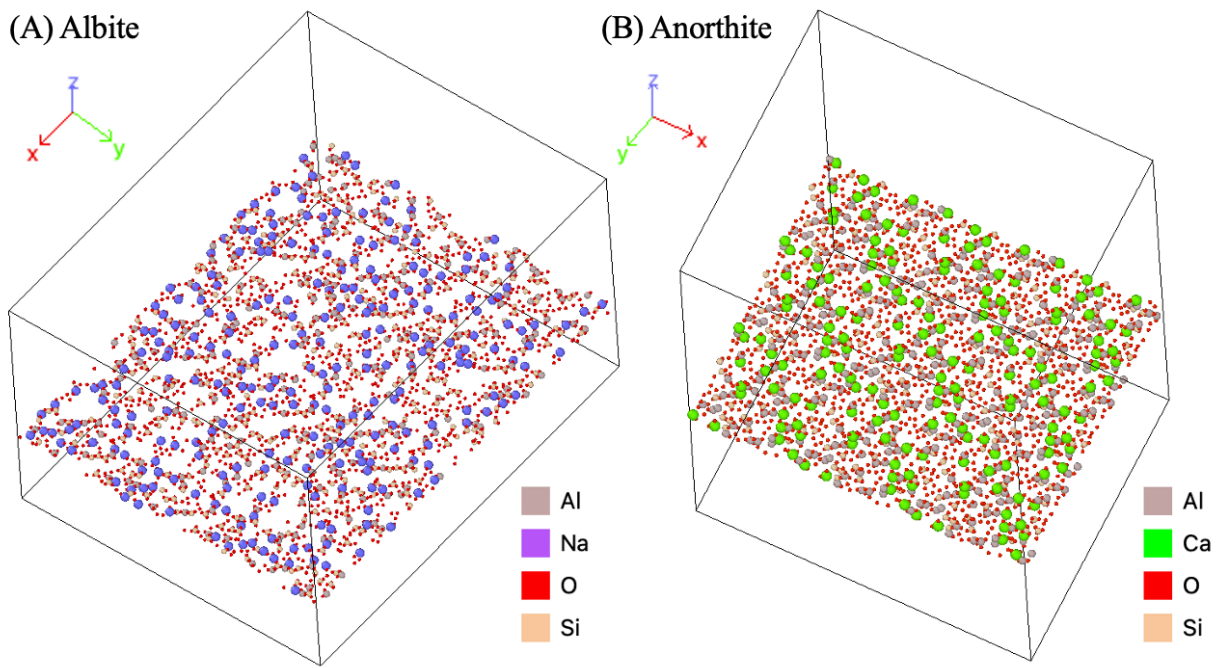

**Figure 2** Albite (A) and anorthite (B) surface atoms only.

The SBE of each identified surface atom was then determined using the same approach as Morrissey et al. (2024b) for crystalline surfaces and previous studies (Yang & Hassanein 2014; Bringuier et al. 2019; Morrissey et al. 2021). This approach involved iteratively applying a velocity perpendicular to the surface. The subsequent position and remaining energy were tracked until the minimum energy needed to remove the surface atom completely from the surface was determined (i.e., no attractive or repulsive forces experienced). The resulting SBEs were then collected into 1 eV bins. SBE distributions of normalized frequencies for each element type were generated by dividing the number of counts for each 1 eV bin by the total number of SBE counts.

## 2.2. BCA Model Simulations

One of the more common current approaches used in planetary science today is the extended version of the BCA Monte Carlo code Transport Range of Ions (TRIM; Ziegler & Biersack 1985) known as SDTrimSP (Mutzke et al. 2019), which can be run in standard (S) or dynamical (D) mode (the latter of which tracks compositional changes in the impacted substrate) using either serial (S) or parallel (P) processing. SDTrimSP is a BCA model used to predict the energy distribution, yield, and angular distribution of atoms ejected from a surface, based on the type, energy, and angle of incoming ions (Biersack & Eckstein 1984; Arredondo et al. 2018; Morrissey et al. 2023; Szabo et al. 2018, 2020). It has demonstrated greater accuracy than TRIM for ion energies typical of the solar wind (Hofsäss et al. 2014).

The set of SBEs derived from the MD simulations were then used as inputs into the BCA SDTrimSP version 7.0 models to study the effect of the SBEs on the predicted sputtering yield. In its current state, SDTrimSP takes the SBE and atomic concentration of each atom type in the overall compound as an input, and cannot accept an SBE distribution as a single input. SBE distributions obtained for each constituent were sorted into 1 eV bins. In SDTrimSP we then defined every bin as a different "atom type" with a unique SBE. The probability for a bin to occur for a given element is equal to its normalized frequency weighted by the atomic fraction of the element in question. Following previous studies using SDTrimSP for SW impacts (Morrissey et al. 2023; Szabo et al. 2020) we considered a cosine distribution of incidence angles with the SBE of each element declared separately via an ISBV =1. Next, in accordance with the recommendation by Möller and Posselt (2002), the experimentally measured bulk densities of albite (2.3 g/cm$^3$) and anorthite (2.5 g/cm$^3$) were obtained by modifying the atomic density of O. Finally, for the cases of 100% 1 keV H and 96% 1 keV H + 4% 4 keV He, 1,000,000 projectiles were launched at the target substrate, while only 100,000 impacts were simulated in the 100% 4 keV He case. In all cases, the target substrate thickness was set to 3,000Å with 150Å layers.

## 3　　　RESULTS AND DISCUSSION
### 3.1　　　SBEs from MD Simulations

The SBE distributions for each element type from albite and anorthite are given in Figure 3. The mean, standard deviation (SD), minimum and maximum SBE values for each element type for amorphous albite and anorthite are given in Tables 1 and 2, respectively, and are compared to the results for their crystalline counterparts presented in Morrissey et al. (2024).

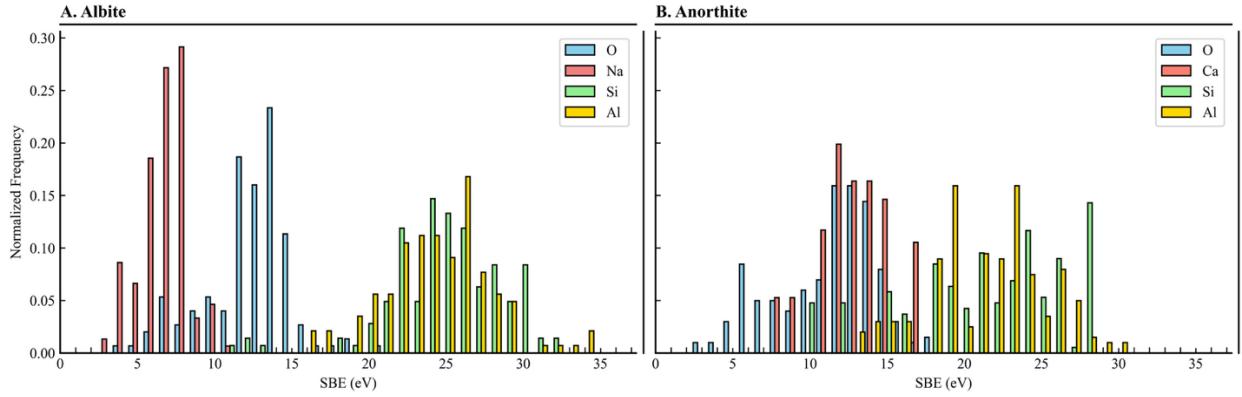

**Figure 3** Frequency distributions for the elemental SBEs of albite (A) and anorthite (B). SBEs are plotted in 1 eV bins.

**Table 1**

MD-Computed Mean SBE, standard deviation (SD), Minimum and Maximum Values for Each Surface Element Type of Amorphous (present study) and Crystalline Albite (L.S. Morrissey et al. 2024)

| Element | Mean SBE (eV) | | SD (eV) | | Minimum (eV) | | Maximum (eV) | |
| --- | --- | --- | --- | --- | --- | --- | --- | --- |
| | Amorph | Crys | Amorph | Crys | Amorph | Crys | Amorph | Crys |
| O | 12.8 | 10.6 | 2.8 | 3.3 | 4.8 | 3.3 | 21.2 | 16.8 |
| Na | 7.3 | 8.2 | 1.5 | 1.7 | 3.6 | 4.1 | 11.6 | 10.7 |
| Si | 25.3 | 26.5 | 3.7 | 6.3 | 11.6 | 12.5 | 32.8 | 35.3 |
| Al | 24.7 | 26.2 | 3.5 | 4.5 | 16 | 19.3 | 34.8 | 32 |

**Table 2**

MD-Computed Mean SBE, SD, Minimum and Maximum Values for Each Surface Element Type of Anorthite and Previously Reported Mean, SD, Minimum and Maximum SBE values for Each Element Type of Crystalline Anorthite (L.S. Morrissey et al., 2024)

| Element | Mean SBE (eV) | | SD (eV) | | Minimum (eV) | | Maximum (eV) | |
| --- | --- | --- | --- | --- | --- | --- | --- | --- |
| | Amorph | Crys | Amorph | Crys | Amorph | Crys | Amorph | Crys |
| O | 11.7 | 10.4 | 3.3 | 2.8 | 3.2 | 5.7 | 18 | 14.5 |
| Ca | 13.3 | 12.7 | 2.2 | 2.3 | 8.2 | 8.1 | 17.4 | 14.8 |
| Si | 21.7 | 22.5 | 5.1 | 7.2 | 10.2 | 15 | 28.6 | 35.5 |
| Al | 21.9 | 21.1 | 3.7 | 7.7 | 13.2 | 13.1 | 30 | 37.6 |

For both crystalline and amorphous albite, Na and O atoms exhibit lower mean SBE values as compared to Si and Al. A similar trend is observed for crystalline and amorphous anorthite, with Ca and O atoms again having lower SBEs than Si and Al. These differences in SBEs can be attributed to the different bond strengths within the mineral. Na and Ca predominantly form ionic bonds with O, which are generally weaker (257 and 464 kJ/mol, respectively) than the covalent Si-O (798 kJ/mol) and Al-O (512 kJ/mol) bonds and thus present a lower SBE (Dean 1999). This makes Na and Ca more easily sputtered or desorbed from the surface compared to Si and Al. In addition, even in an amorphous surface, the Si and Al atoms are still relatively well coordinated (forming more O bonds) in the silicate network, requiring more energy to remove them. In contrast, O surface atoms may be undercoordinated or in weaker bonding configurations compared to those in the bulk, making them easier to remove.

When comparing the mean and SD of the SBEs for crystalline and amorphous substrates there are minimal differences for both albite and anorthite. This can be explained by the similar local bond types being formed for each element, regardless of whether the surface is crystalline or amorphous. For example, whether Na is contained in a long-range well-ordered crystalline substrate or a disordered amorphous substrate it still favors ionic bonds with near-neighbour O. However, there are significant differences in the distribution of these SBEs for crystalline and amorphous surfaces. In crystalline surfaces, SBEs exhibit a discretized range for each element type because the atoms occupy well-defined, repeating positions, leading to a finite set of binding sites. In contrast, amorphous surfaces produce a broad range of SBEs due to their disordered structure, which allows for a continuous variation in bonding configurations. In addition, there are no notable trends among the minimum and maximum SBE values when comparing the amorphous surfaces to their crystalline counterparts. We suspect this is due to the melting procedure occurring during amorphization. This has a two-fold effect. First, for undercoordinated crystalline atoms with low SBEs the amorphization allows them to post into lower and better coordinated positions, thus increasing their SBE. Second, for highly coordinated atoms with initially high SBEs, the local disorder introduced during amorphization may disrupt their coordination environment, leading to a decrease in their SBE. As a result, amorphization can both raise and lower SBE values depending on the local atomic configuration, leading to no consistent trend when comparing the amorphous and crystalline surfaces.

Further investigation of the minimum and maximum SBE values for each element in the amorphous albite substrate was performed using OVITO to visualize the bonding environments responsible for these extremes, as shown in Figure 4.

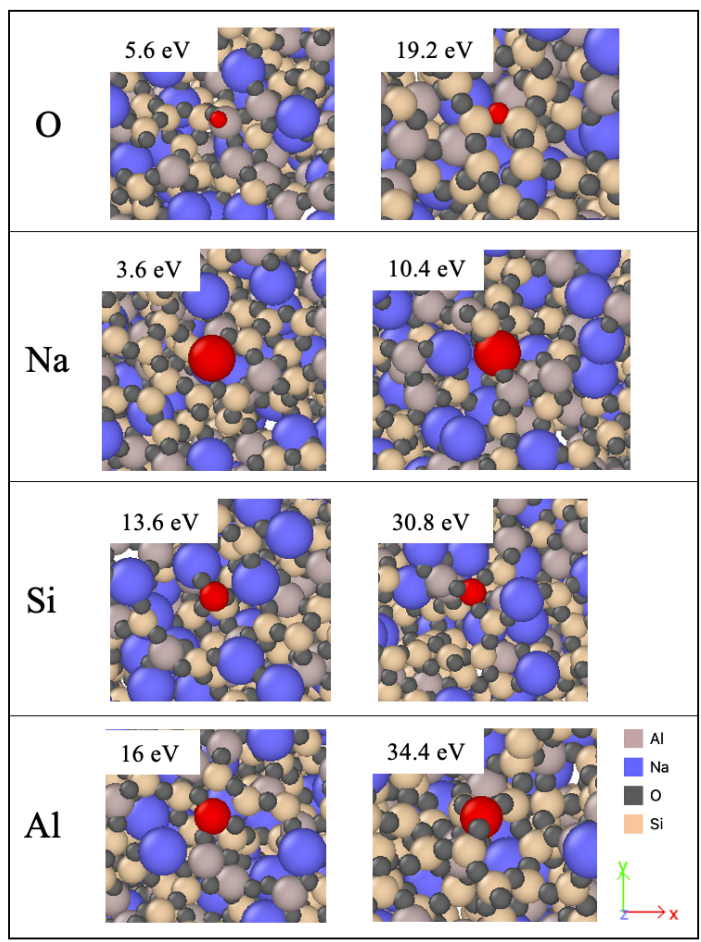

**Figure 4**. Visualization of local bonding environments for atoms with near minimum and near maximum SBE values for each element in amorphous albite.

It is evident that the minimum (low) SBE values in the distributions for each element type arise from atoms situated at local surface peaks, where they are surrounded by fewer neighbouring atoms. In contrast, the maximum (high) SBE values arise from atoms in local surface valleys, where they are more strongly bound by surrounding neighbours that must be overcome for ejection. Essentially, on amorphous surfaces with many peaks and valleys, atomic-scale roughness dominates the SBE. Further work, beyond the scope of this paper, is required to investigate the precise bond configurations that give rise to the binding energies observed at each local configuration.

## 3.2     Effect on Predicted Sputtering Behaviour

SDTrimSP simulations were conducted to predict the sputtering behaviour of the amorphous albite and anorthite surfaces using the MD-derived SBEs. The elemental sputtering yields and total substrate sputtering yields for amorphous albite and anorthite are presented and compared to the crystalline albite and anorthite results reported by Morrissey et al. (2024) in Tables 3 and 4, respectively.

**Table 3**

Summed Elemental Sputtering Yields and Total Substrate Sputtering Yields from Amorphous Albite for the Three Different Impacting Ion Cases and Previously Reported Counterpart Values from Crystalline Albite (L.S. Morrissey et al., 2024)

| | | | Sputtering Yield (atoms/ion) | | | | | |
| --- | --- | --- | --- | --- | --- | --- | --- | --- |
| | SBE (eV) | | 100% H | | 100% He | | 96% H + 4% He | |
| Element | Amorph | Crys | Amorph | Crys | Amorph | Crys | Amorph | Crys |
| O | 12.8 | 10.6 | 9.12E-03 | 1.23E-02 | 8.27E-02 | 1.02E-01 | 1.20E-02 | 1.59E-02 |
| Na | 7.3 | 8.2 | 2.01E-03 | 1.79E-03 | 1.49E-02 | 1.43E-02 | 2.56E-03 | 2.30E-03 |
| Si | 25.3 | 26.5 | 5.39E-04 | 5.12E-04 | 1.43E-02 | 1.43E-02 | 1.10E-03 | 1.06E-03 |
| Al | 24.7 | 26.2 | 1.88E-04 | 1.55E-04 | 4.78E-03 | 4.59E-03 | 3.80E-04 | 3.29E-04 |
| Total Sputtering Yield | | | 1.19E-02 | 1.48E-02 | 1.17E-01 | 1.35E-01 | 1.61E-02 | 1.96E-02 |

**Table 4**

Summed Elemental Sputtering Yields and Total Substrate Sputtering Yields from Amorphous Anorthite for the Three Different Impacting Ion Cases and Previously Reported Counterpart Values from Crystalline Anorthite (L.S. Morrissey et al., 2024)

| | | | Sputtering Yield (atoms/ion) | | | | | |
| --- | --- | --- | --- | --- | --- | --- | --- | --- |
| | SBE (eV) | | 100% H | | 100% He | | 96% H + 4% He | |
| Element | Amorph | Crys | Amorph | Crys | Amorph | Crys | Amorph | Crys |
| O | 11.7 | 10.4 | 1.12E-02 | 1.21E-02 | 9.52E-02 | 1.03E-01 | 1.46E-02 | 1.55E-02 |
| Ca | 13.3 | 12.7 | 4.86E-04 | 4.39E-04 | 7.59E-03 | 7.68E-03 | 7.34E-04 | 7.44E-04 |
| Si | 21.7 | 22.5 | 6.06E-04 | 6.29E-04 | 1.16E-02 | 1.22E-02 | 1.00E-03 | 9.44E-04 |
| Al | 21.9 | 21.1 | 5.34E-04 | 6.26E-04 | 1.11E-02 | 1.20E-02 | 9.74E-04 | 1.06E-03 |
| Total Sputtering Yield | | | 1.28E-02 | 1.38E-02 | 1.25E-01 | 1.35E-01 | 1.73E-02 | 1.83E-02 |

Firstly, many of the same trends and insights reported by Morrissey et al. (2024) for crystalline surfaces (i.e., the preferential sputtering of atoms with lower SBEs and the non-uniform increase in sputtering yield across element types when the solar wind includes 4% He) are also observed for amorphous surfaces. This suggests that these behaviors are not affected by the considered changes in structural arrangement of the minerals. As shown in Table 3, amorphous albite exhibits slightly higher sputtering yields than crystalline for all element types except O, with percent differences remaining below 20%. The mean amorphous SBE values for these elements are also slightly lower than the crystalline mean SBE values, suggesting that the more loosely bound atoms are contributing to the higher sputtering yield. In contrast, the O sputtering yield decreases for amorphous albite when compared to crystalline albite, with percent differences above 20%. The mean SBE value for O in amorphous albite is higher than that for O in crystalline albite, suggesting that the more tightly bound atoms are reducing the sputtering yield. The total sputtering yield for amorphous albite is lower than that of crystalline albite but agrees within 20% for all impacting ion cases as presented in Table 3.

Similar to albite, there are again minimal differences in the elemental and total yields for amorphous and crystalline anorthite, with percent differences less than 10%. However, the majority of the elemental sputtering yields decreased for amorphous anorthite when compared to the yields for crystalline anorthite as shown in Table 4. That is, only Ca for the 100% H impacting ion case and Si for the 4% He + 96% H impacting ion case increased very slightly, less than 10%. In all cases besides Si the mean SBE values for each elemental type increased in amorphous anorthite when compared to crystalline, indicating that the tightly bound atoms are reducing the sputtering yields. These minimal decreases in elemental sputtering yields translate to a slightly reduced total sputtering yield of amorphous anorthite when compared to those of crystalline anorthite.

Morrissey et al. (2024) also demonstrated that a yield-weighted single SBE can be used in the Thompson theory to approximate the overlapping energy distributions instead of considering all SBEs. The same approach can be used here for the resulting SBE distributions and these yield-weighted single elemental SBEs are provided in Table 5. These values can be used in the Thompson model to predict ejecta energy distributions, producing similar energy distributions to those presented in Morrissey et al. 2024, since these amorphous yield-weighted elemental SBEs are similar to their crystalline counterparts.

Table 5
Yield-Weighted Elemental SBEs for Amorphous Anorthite for the Three Different Impacting Ion Cases

| Mineral | Element | Yield-Weighted SBE (eV) | | |
| --- | --- | --- | --- | --- |
| | | 100% H | 100% He | 96% H + 4% He |
| Albite | Na | 6.4 | 6.6 | 6.4 |
| | Al | 23.0 | 23.8 | 23.5 |
| | Si | 22.8 | 24.2 | 23.5 |
| | O | 11.3 | 11.8 | 11.4 |
| Anorthite | Ca | 12.1 | 12.6 | 12.2 |
| | Al | 19.9 | 20.9 | 20.2 |
| | Si | 17.8 | 20.0 | 19.0 |
| | O | 9.6 | 10.3 | 9.8 |

3.3     Implications and Comparison to Previous Studies

While direct comparative studies on the sputtering behaviour of crystalline versus amorphous plagioclase feldspars are limited, previous experiments and theoretical studies disagree on the effects of atomic arrangement/crystallinity on the yield. In a compilation of theoretical and experimental studies on a range of different substrates, Behrisch and Eckstein (2007) suggested that sputtering yields from randomly oriented polycrystalline substrates can be approximated to a first order by the yields for an amorphous substrate. This was also validated by comparison of crystalline, polycrystalline and amorphous sputtering yields of copper using MD modelling (Morrissey et al. 2021). In contrast, Schlueter et al. 2020 showed that the sputtering yield of a polycrystalline metal is discrepant from its amorphous counterpart when exposed to heavy keV ions (30 keV Ge), with the yield increasing by a factor of two in the polycrystalline case. They attribute this difference to the unique linear collision sequences that occur in each substrate type. Here we consider significantly lighter and lower energy impactors. We see agreement in all cases within 20% and attribute this to the similar bond types, and thus SBEs, found in both cases. While the distributions and discretized nature of the SBEs are different, Na/Ca/Si/Al atoms are all still bonding with O and thus producing similar values. Moreover, during amorphization the high temperatures produced to allow for melting can lead to the flattening of an atomically rough surface and the removal of some of the most undercoordinated and loosely bound positions found in a crystalline lattice. The findings herein suggest that crystalline samples can serve as a good approximation for amorphous samples when studying sputtering by the solar wind.

**4         CONCLUSIONS**

The results presented here have built on the work previously completely by our group by considering amorphous substrates. We have used a combined MD/BCA approach to obtain SBEs and sputtering yields for amorphous albite and anorthite, allowing for comparison to their crystalline counterparts. Findings from this study support the conclusion that amorphization of crystalline albite and anorthite results in wider and smoother atomic SBE distributions due to the uniquely randomized atomic positions. In a crystal, atoms occupy unique repeating positions resulting in well discretized SBEs, whereas in an amorphous substrate randomness dominates and these peaks in the distribution are lost due to many slightly different configurations. However, since the same bonds between atoms are present in both the crystalline and amorphous minerals the mean atomic SBEs are very similar and the effects of amorphization on sputtering yield are limited when compared to polycrystalline targets. It is important to note that these studies do not account for the dynamic roughness that may develop in crystalline or amorphous minerals with increased fluence. This damage process occurs when an atom is sputtered and thus leaves behind undercoordinated surface atoms with a potentially modified SBE. The effect of this process in producing a *dynamic SBE* remains unknown and will be considered in a future study. Such effects may be important for refining exosphere models of the Moon and Mercury.

**Acknowledgements**

This research was supported by the International Space Science Institute (ISSI) in Bern, through ISSI International Team project #24-616 "Multi-scale Understanding of Surface-Exosphere Connections (MUSEC). A.R., B.A.C.-G., A.G.

and L.S.M. were supported in part by the NSERC Discovery Grant and the CSA Research Opportunities in Space Sciences program awards. A.R. would like to acknowledge Daniel W. Savin from the Columbia Astrophysics Laboratory at Columbia University for the stimulating discussions and for the support in the preparation of this manuscript.

## 5 REFERENCES

Arredondo, R., Oberkofler, M., Schwarz-Selinger, T., von Toussaint, U., Burwitz, V.V., Mutzke, A., Vassallo, E., & Pedroni, M. (2018). Angle-dependent sputter yield measurements of keV D ions on W and Fe and comparison with SDTrimSP and SDTrimSP-3D. *Nuclear Materials and Energy*, *18*, 72-76.

Behrisch, R., & Eckstein, W. (2007). *Sputtering by Particle Bombardment: Experiments and Computer Calculations from Threshold to MeV Energies*. Springer.

Bennett, C.J., Pirim, C., & Orlando, T.M. (2013). Space-Weathering of Solar System Bodies: A Laboratory Perspective. *Chemical Reviews*, *113*(12), 9086-9150.

Biersack J.P., & Eckstein, W. (1984). Sputtering Studies with the Monte Carlo Program TRIM.SP. *Applied Physics*.

Bringuier, S., Abrams, T., Guterl, J., et al. (2019). Atomic insight into concurrent He, D, and T sputtering and near-surface implantation of 3D-SiC crystallographic surfaces. *Nuclear Materials and Energy*, *19*, 1-6.

Brötzner, J., Biber, H., Szabo, P.S., Jäggi, N., Fuchs, L., Nenning, A., Fellinger, M., Nagy, G., Pitthan, E., Primetzhofer, D., Mutzke, A., Wilhelm, R.A., Wurz, P., Galli, A., & Aumayr, F. (2025). Solar wind erosion of lunar regolith is suppressed by surface morphology and regolith properties. *Communications Earth & Environment*, *6*, 560.

Cui, X., Ringer, S.P., Wang, G., & Stachurski, Z.H. (2019). What should the density of amorphous solids be? *The Journal of Chemical Physics*, *151*.

Dal Bó, M., Cantavella, V., Sánchez, E., Hotza, D., & Gilabert, F.A. (2013). Fracture toughness and temperature dependence of Young's modulus of a sintered albite glass. *Journal of Non-Crystalline Solids*, 70-76.

Dean, J.A. (Ed.). (1999). *Lange's handbook of chemistry* (15th ed., Section 4: Properties of atoms, radicals, and bonds). McGraw-Hill.

Domingue, D.L., Chapman, C.R., Killen, R.M., Zurbuchen, T.H., Gilbert, J.A., Sarantos, M., et al. (2014). Mercury's Weather-Beaten Surface: Understanding Mercury in the Context of Lunar and Asteroidal Space Weathering Studies. *Space Science Reviews*, *181*, 121-214.

Fogarty, J.C., Aktulga, H.M., Grama, A.Y., Van Duin, A.C.-T., & Pandit, S.A. (2010). A reactive molecular dynamics simulation of the silica-water interface. *The Journal of Chemical Physics, 132*(17).

Grossman, J.J., Ryan, J.A., Mukherjee, N.R., & Wegner, M.W. (1970). Surface Properties of Lunar Samples. *Science*, *167*(3918), 743-745.

Hapke, B. (2001). Space weathering from Mercury to the asteroid belt. *Journal of Geophysical Research*, *106*(5), 10039-10074.

Hofsäss, H., Zhang, K., & Mutzke, A. (2014). Simulation of ion beam sputtering with SDTrimSP, TRIDYN, and SRIM. *Applied Surface Science*, *310*(15), 134-141.

Jain, A., Ong, S.P., Hautier, G., Chen, W., Richards, W.D., Dacek, S., Cholia, A., Gunter, D., Skinner, D., Ceder, G., & Persson, K.A. (2013). The Materials Project: A materials genome approach to accelerating materials innovation. *APL Materials*, *1*.

Keller, L.P., & McKay, D.S. (1991). The origin of amorphous rims on lunar plagioclase grains: Solar wind damage or vapor condensates. *54th Annual Meet. Meteorit. Soc.*, *766*.

Keller, L.P., & McKay, D.S. (1994). The contribution of vapor deposition to amorphous rims on lunar soil grains. *Meteoritics*, *29*(4).

Keller, L.P., & McKay, D.S. (1997). The nature and origin of rims on lunar soil grains. *Geochimica et Cosmochimica Acta*, *61*(11), 2331-2341.

Kelly, R. (1986). The surface binding energy in slow collisional sputtering. *Nuclear Instruments and Methods in Physics Research Section B: Beam Interactions with Materials and Atoms, 18*(1-6), 388-398.

Killen, R.M., Morrissey, L.S., Burger, M.H., Vervack, R.J., Tucker, O.J., & Savin, D.W. (2022). The Influence of Surface Binding Energy on Sputtering in Models of the Sodium Exosphere of Mercury. *The Planetary Science Journal*, *3*(6).

Lyngdoh, G.A., Kumar, R., Krishnan, N.M.A., & Das, S. (2019). Realistic atomic structure of fly ash-based geopolymer gels: Insights from molecular dynamics simulations. *The Journal of Chemical Physics*, *151*.

Mayanovic, R.A., Anderson, A.J., Romine, D., & Benmore, C.J. (2023). Insights on the dissolution of water in an albite melt at high pressures and temperatures from a direct structural analysis. *Scientific Reports*, *13*(1), 1–10.

Morrissey, L.S., Bringuier, S., Bu, C., Burger, M.H., Dong, C., Ebel, D.S., Harlow, G.E., Huang, Z., Killen, R.M., Leblanc, F., Ricketts, A., Tucker, O.J., & Savin, D.W. (2024). Solar Wind Ion Sputtering from Airless Planetary Bodies: New Insights into the Surface Binding Energies for Elements in Plagioclase Feldspars. *The Planetary Science Journal*, *5*, 272.

Morrissey, L.S., Pratt, D., Farrell, W.M., Tucker, O.J., Nakhla, S., & Killen, R.M. (2022a). Simulating the diffusion of hydrogen in amorphous silicates: A "jumping" migration process and its implications for solar wind implanted lunar volatiles. *Icarus*, *379*, p. 114979.

Morrissey, L.S., Schaible, M.J., Tucker, O.J., Szabo, P.S., Bacon, G., Killen, R.M., & Savin, D.W. (2023). Establishing a Best Practice for SDTrimSP Simulations of Solar Wind Ion Sputtering. *The Planetary Science Journal*, *4*(67).

Morrissey, L.S., Tucker, O.J., Killen, R.M., Nakhla, S., & Savin, D.W. (2021). Sputtering of surfaces by ion irradiation: A comparison of molecular dynamics and binary collision approximation models to laboratory measurements. *Journal of Applied Physics*, *130*(1).

Morrissey, L.S., Tucker, O.J., Killen, R.M., Nakhla, S., & Savin, D.W. (2022b). Solar Wind Ion Sputtering of Sodium from Silicates Using Molecular Dynamics Calculations of Surface Binding Energies. *The Astrophysical Journal Letters*, *925*(6).

Mutzke, A., Schneider, R., Eckstein, W., Dohmen, R., Schmid, K., von Toussaint, U., & Bandelow, G. (2019). SDTrimSP Version 6.00 IPP 2019-02, *Max-Planck-Institut für Plasmaphysik*.

Möller W., & Posselt M. (2002). TRIDYN_FZR User Manual. *Forschungszentrum Rossendorf*.

Pallini, A., Bertani, M., Rustichelli, D., Ziebarth, B., Mannstadt, W., & Pedone, A. (2023). Comparison of five empirical potential models for aluminosilicate systems: Albite and anorthite as test cases. *Journal of Non-Crystalline Solids*, *615*, 122426.

Pieters, C.M., & Noble, S.K. (2016). Space Weathering on Airless Bodies. *Journal of Geophysical Research: Planets*, *121*(10), 1865-1884.

Pitman, M.C., & Van Duin, A.C.-T. (2012). Dynamics of Confined Reactive Water in Smectite Clay-Zeolite Composites. *Journal of the American Chemical Society*, *134*(6), 3042-3053.

Plimpton, S.J. (1995). Fast Parallel Algorithms for Short-Range Molecular Dynamics. *Journal of Computational Physics*, *117*(1), 1-19.

Schlueter, K., Nordlund, K., Hobler, G., Balden, M., Grandberg, F., Flinck, O., da Silva, T.F., & Neu, R. (2020). Absence of a crystal direction regime in which sputtering corresponds to amorphous material. *Physical Review Letters*, *125*(22), 225502.

Stukowski, A. (2010). Visualization and analysis of atomistic simulation data with OVITO – the Open Visualization Tool. *Modelling and Simulation in Materials Science and Engineering*, *18*(1).

Szabo, P.S., Biber, H., Jäggi, N., Brenner, M., Weichselbaum, D., Niggas, A., Stadlmayr, R., Primetzhofer, D., Nenning, A., & Mutzke, A. (2020). Dynamic potential sputtering of lunar analog material by solar wind ions. *The Astrophysical Journal*, *891*(1).

Szabo, P.S., Chiba, R., Biber, H., Stadlmayr, R., Berger, B.M., Mayer, D., Mutzke, A., Doppler, M., Sauer, M., Appenroth, J., Fleig, J., et al. (2018). Solar wind sputtering of wollastonite as a lunar analogue material – Comparisons between experiment and simulations. *Icarus*, *314*, 98-105.

Szabo, P.S., Poppe, A.R., Mutzke, A., Liuzzo, L., & Carberry Mogen, S.R. (2024). Backscattering of Ions Impacting Ganymede's Surface as a Source for Energetic Neutral Atoms. *The Astrophysical Journal Letters, 963*(1), 8.

Thompson, A.P., Aktulga, H.M., Berger, R., et al. (2022). LAMMPS – a flexible simulation tool for particle-based materials modeling at the atomic, meso, and continuum scales. *Computer Physics Communications*, *271*.

Van Duin, A.C.-T., Dasgupta, S., Lorant, F., & Goddard, W.A. (2001). ReaxFF: A Reactive Force Field for Hydrocarbons. *Journal of Physical Chemistry A*, *105*(41), 9396-9409.

Yang, X., & Hassanein A. (2014). Atomic scale calculations of tungsten surface binding energy and beryllium-induced tungsten sputtering. *Applied Surface Science*, *293*, 187-190.

Yu, Y., Wang, B., Wang, M., Sant, G., & Bauchy, M. (2017). Reactive Molecular Dynamics Simulations of Sodium Silicate Glasses — Toward an Improved Understanding of the Structure. *International Journal of Applied Glass Science*, *8*(3), 276–284.

Ziegler, J.F., & Biersack, J.P. (1985). The Stopping and Range of Ions in Matter. In D.A. Bromley (Ed.), *Treatise on Heavy-ion Science* (Vol. 6, pp. 93-129). Springer.

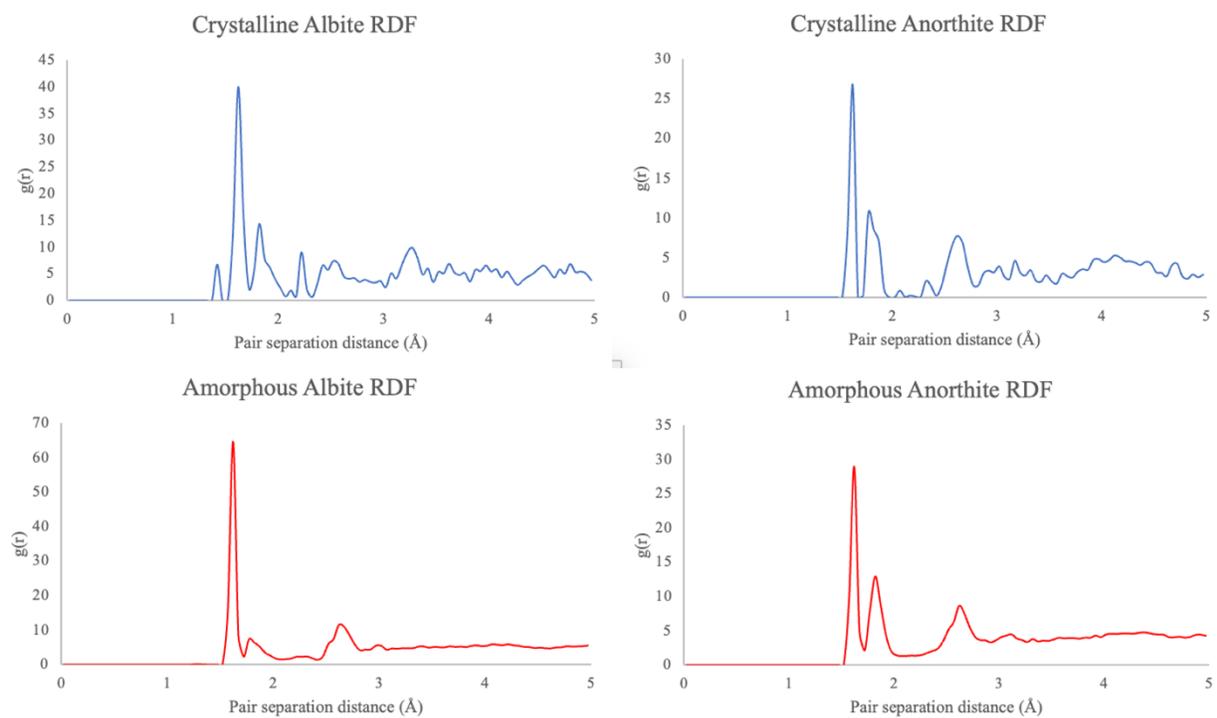

**Figure S1.** Radial distribution function (RDF) analyses performed in OVITO for the crystalline and amorphous albite and anorthite structures, showing the variation in atomic ordering between the two phases.